\documentclass[12pt]{article}
\usepackage{graphics}
\usepackage{epsfig}
\usepackage{subfigure}
\usepackage{graphics}
\usepackage{amsmath}
\usepackage{amsfonts}
\usepackage{amssymb}
\usepackage{url}
\usepackage{hyperref}

\newcommand{\be}{\begin{equation}}
\newcommand{\ee}{\end{equation}}
\newcommand{\bea}{\begin{eqnarray}}
\newcommand{\eea}{\end{eqnarray}}

\newcommand{\GeV}{~\mathrm{GeV}}
\newcommand{\TeV}{~\mathrm{TeV}}
\newcommand{\zp}{Z^{\prime}}

\def\nn{\hspace{2mm}}
\def\sss{\scriptscriptstyle}
\def\GeV{\mbox{\rm GeV}}

\def\TeV{\mbox{\rm TeV}}

\def\abs#1{\left| #1\right|} 

\begin{document}
%%%%%%%%%%%%%%%%%%%%%%%%%%%%
\begin{titlepage}
%%%%% PREPRINT NUMBERS %%%%%%
\begin{flushright}
KEK-TH-1247
\end{flushright}
%%%%%%%%%%%%%%%%%%%%%%%%%%%%%%
\vspace{4\baselineskip}
%%%%%%%%%%%%%%%%%%% TITLE %%%%%%%%%%%%%%%%%%
\begin{center}
{\Large\bf 
Radiative $B-L$ symmetry breaking \\
and the $Z'$ mediated SUSY breaking
}
\end{center}
%%%%%%%%%%%%%%%% AUTHORS %%%%%%%%%%%%%%%%%%%%%%%
\vspace{1cm}
\begin{center}
{\large
Tatsuru Kikuchi$^{a}$
\footnote{\tt E-mail:tatsuru@post.kek.jp}
and Takayuki Kubo$^{a,b}$
\footnote{\tt E-mail:kubotaka@post.kek.jp}
}
\end{center}
%%%%%%%%%%%%%%%%%%%%%%% AFFILIATION %%%%%%%%%%%%
\vspace{0.2cm}
\begin{center}
${}^{a}$ 
{\small \it Theory Division, KEK,
Oho 1-1, Tsukuba, Ibaraki, 305-0801, Japan}\\
${}^{b}$ 
{\small \it The Graduate University for Advanced Studies,
Oho 1-1, Tsukuba, Ibaraki, 305-0801, Japan}\\
\medskip
\vskip 5mm
\end{center}
\vskip 5mm
\begin{abstract}
We explore a mechanism of radiative $B-L$ symmetry breaking
in analogous to the radiative electroweak symmetry breaking.
The breaking scale of $B-L$ symmetry is related to the neutrino masses
through the see-saw mechanism.
Once we incorporate the ${\rm U}(1)_{\rm B-L}$ gauge symmetry in SUSY models,
the ${\rm U}(1)_{\rm B-L}$ gaugino, $\tilde{Z}_{B-L}$ appears, and it can mediate 
the SUSY breaking (Z-prime mediated SUSY breaking) at around the scale of $10^6$ GeV.
Then we find a links between the neutrino mass (more precisly the see-saw or $B-L$ scale
of order $10^{6}$ GeV) and the Z-prime mediated SUSY breaking scale.
It is also very interesting that the gluino at the weak scale becomes relatively light,
and almost compressed mass spectra for the gaugino sector can be
realized in this scenario, which is very interesting in scope of the LHC.
\end{abstract}
\end{titlepage}
%%%%%%%%%%%%%%%%%%%%%%%%%%%%%%%%%%%%%%%%%%%%%%%%
\section{Introduction}
Based on the experimental data, now the evidence of the neutrino masses and flavor mixings 
are almost established, and this is also the evidence of new physics beyond the standard model. 
Interestingly, the neutrino mass and mixing properties have been revealed to be very different 
from those of the other fermions, namely, neutrino masses are very small and the flavor mixing
angles are very large. A new physics must explain them.

Supersymmetry (SUSY) extension of the Standard Model (SM) is one of the attractive candidates 
for new physics \cite{SUSY}. This is one of the most promising way to solve the gauge hierarchy problem 
in the standard model. The experimental data support the unification of 
the three gauge couplings at the grand unified theory (GUT) scale 
$M_{\rm GUT} \sim 2 \times 10^{16}$ GeV 
with the particle contents of the minimal supersymmetric standard model (MSSM) 
\cite{unification1, unification2}. 
If we use the see-saw mechanism \cite{seesaw}, it can naturally explain 
the lightness of the neutrinos.

The experimental data suggests that the see-saw scale is much lower than
the Planck scale or even the GUT scale. It is therefore natural to think 
the scale is related to the breaking of some symmetry. 
The simplest symmetry is the $B-L$ symmetry. 
In principle, the $B-L$ symmetry can be a global or a local symmetry. 
If we take it to be a global symmetry, its spontaneous breaking leads to the pseudo Nambu-Goldstone
boson, majoron. Since several experiments give severe constraints on the majoron,
it is natural to make it local gauge symmetry if we consider a higher ranked GUT such as
${\rm SO}(10)$ \footnote{For the group theoretical aspects of {\rm SO}(10), see for example,
\cite{FIKMO}.}.
The spontaneous breaking of $B-L$ symmetry can be exploited by developing
the vacuum expectation value (VEV) of a scalar multiplet $\Delta_1$
which carries $B-L=-2$. For the anomaly cancellation and to keep the low-energy
supersymmetry, its counterpart $\Delta_2$ that has $B-L=+2$ has to be included
into a theory. After the spontaneous breaking of this $B-L$ symmetry, it leads to
a massive gauge boson, $Z_{B-L}$.

On the other hand, one of the most attractive features of the Minimal Supersymmetric extension
of the Standard Model (MSSM), is the fact that it provides a mechanism for radiative breaking of
the electroweak gauge ${\rm SU}(2)_L \times {\rm U}(1)_Y$ symmetry 
\cite{Inoue:1982pi, Ibanez:1982fr, Ibanez:1983wi, Alvarez-Gaume:1983gj,  Ellis:1982wr, Ellis:1983bp}. 
The essential point for this mechanism is that the presence of the large top Yukawa coupling, which
can dictate the Higgs mass squared driven to be negative at the weak scale.
It is known that the radiative electroweak symmetry breaking (RESB) can take place
if the top Yukawa coupling is large, such that $60\, {\rm GeV} \lesssim M_t \lesssim 200\, {\rm GeV}$, 
with the upper bound coming from the requirement that it remains in the perturbative range up to
the GUT scale. It is interesting that the observed top quark mass found at the Tevatron 
was indeed at around $M_t = 175$ GeV.

Then it is natural to think about the possibility to break ${\rm U}(1)_{\rm B-L}$ symmetry through
the radiative corrections to the soft mass squared which is responsible for the VEV of
the ${\rm U}(1)_{\rm B-L}$ breaking in analogous to the case of RESB in the MSSM.
Here we explore such a possibility by considering the renormalization group equations (RGEs)
of the soft mass terms for the $B-L$ breaking sector.
Our resultant $B-L$ breaking scale is found to be around $v_{B-L} \simeq10^{5}$ GeV, that is
in a  sense quite appealing if we consider to incorporate the thermal leptogenesis scenario \cite{fukugita}
in SUSY models because the gravitino problem \cite{weinberg, Khlopov} put a severe constraint
on the reheating temperature as $T_R \lesssim 10^6$ GeV
for the gravitino mass of order $m_{3/2} \lesssim 100$ GeV \cite{kawasaki}.
 
Once we incorporate the ${\rm U}(1)_{\rm B-L}$ gauge symmetry in SUSY models,
an extra ${\rm U}(1)$ gaugino, $\tilde{Z}_{B-L}$ appears in addition to the extra gauge boson $Z_{B-L}$. 
It has recently been noticed that if there exist such an extra gaugino, it can mediate a SUSY breaking
so as to induce the gaugino masses for each SM gauge group at the two loop level,
while the scalar soft masses are generated at the one loop level \cite{ZMSB1, ZMSB2}
\footnote{The similar idea has also been suggested in \cite{Dobrescu:1997qc}.}.
The Z-prime mediated SUSY breaking is basically to use an extra ${\rm U}(1)^\prime$ vector multiplet
as a field which communicates a SUSY breaking source with the visible sector. 
This setup is much more appealing and economical than the gauge-mediated
SUSY breaking.
In this mediation mechanism, it is not necessary to introduce some additional
sector as a 'messenger field', that can be implemented into a theory just as a gauge
multiplet associated with an extra ${\rm U}(1)^\prime$ gauge symmetry.
We take such an extra ${\rm U}(1)^\prime$ as a ${\rm U}(1)_{\rm B-L}$ symmetry,
and then, we can identify the messenger scale as the scale of $B-L$ symmetry
breaking scale.

This paper is organized as follows. 
In section 2 we give an explicit model having $B-L$ symmetry.
In section 3, we explain the model of $Z_{B-L}$ mediated SUSY breaking
mechanism, and apply it to the case of $B-L$ gauge symmetry.
In section 4, some numerical analysis is performed to show some example
SUSY mass spectra which is characteristic for this specific SUSY breaking
mechanism.
The last section is devoted for summary and discussions.

\section{Radiative B-L breaking}
 The interactions between Higgs and matter superfields are
described by the superpotential %
\begin{eqnarray}%
W &=& (Y_u)_{ij} U^c_i Q_j  H_u + (Y_d)_{ij} D^c_i Q_i  H_d 
+ (Y_e)_{ij} E^c_i L_j H_d 
\nonumber\\
&+& \mu H_d H_u \nonumber\\
&+& (Y_\nu)_{ij} N^c_i L_j H_u+ f_{ij} \Delta_1 N^c_i N^c_j 
\nonumber\\
&+& \mu' \Delta_1 \Delta_2 \;,
\label{superpot}
\end{eqnarray}
where the indices $i$, $j$ run over three generations, $H_u$ and $H_d$ 
denote the up-type and down-type MSSM Higgs doublets, respectively.

After developing the VEV of the $B-L$ breaking field, 
$\left<\Delta_1 \right> = v_{B-L}$,
the right-handed neutrino obtains the Majorana mass as $M_N = f v_{B-L}$.
And it gives a light neutrino mass through the see-saw mechanism as follows:
$M_\nu = m_D M_N^{-1} m_D^T$, where $m_D = Y_\nu v~(v=174\,{\rm GeV})$ 
is the Dirac neutrino mass matrix.

The soft SUSY-breaking terms which is added to the MSSM soft mass terms are given by
\begin{eqnarray}
 -  \Delta{\cal L}_{\rm soft} 
&=& ( m^2_N)_{ij} \tilde{N}_i^{\dagger} \tilde{N}_j 
+m_{\Delta_1}^2 |\Delta_1|^2 
+ m_{\Delta_2}^2 |\Delta_2|^2
+ \left((A_{\nu})_{ij} \tilde{N}_i^{\dagger}  \tilde{\ell}_j H_u  + h.c. \right)
\nonumber\\
&+& (A_f)_{ij} \Delta_1 \tilde{N}_i \tilde{N}_j  +h.c.
\nonumber\\\
&+& \frac{1}{2} M_{\tilde{Z}_{B-L}} \tilde{Z}_{B-L}  \tilde{Z}_{B-L}  +h.c. 
\label{softterms} 
\end{eqnarray}
From Eqs. (\ref{superpot}) and (\ref{softterms}),
the scalar potential relevant for the $B-L$ breaking sector can be written as
\bea
V(\Delta_1,\Delta_2)
&=& \left( |\mu'|^2 + m_{\Delta_1}^2 \right) |\Delta_1|^2 
+ \left( |\mu'|^2 + m_{\Delta_2}^2 \right) |\Delta_2|^2 
\nonumber\\
&+& \frac{1}{2} g_{B-L}^2 \left(|\Delta_1|^2 - |\Delta_2|^2 \right)^2 \;,
\eea
where we have neglected the Yukawa coupling contributions to the scalar potential.

The minimalization condition of this potential leads to
\bea
\frac{\partial V}{\partial \Delta_1^\dag} &=& 
\left[\left( |\mu'|^2 + m_{\Delta_1}^2 \right) 
 +  \frac{1}{2} g_{B-L}^2 |\Delta_1|^2 \right] \Delta_1 =0 \;,
\nonumber\\
\frac{\partial V}{\partial \Delta_2^\dag} &=& 
\left[\left( |\mu'|^2 + m_{\Delta_2}^2 \right) 
+ \frac{1}{2} g_{B-L}^2 |\Delta_2|^2  \right] \Delta_2 =0 \;.
\eea
Therefore, the VEV of the $B-L$ breaking field $\Delta_1$ is determined to be
\be
|\left< \Delta_1 \right>|^2 =
- \frac{2}{g_{B-L}^2} \left( |\mu'|^2 + m_{\Delta_1}^2 \right)  \;.
\ee

\section{Z-prime mediation of SUSY breaking}
Here we give a brief review of the Z-prime mediation of SUSY breaking \cite{ZMSB1, ZMSB2}
by discussing the pattern of the soft SUSY breaking parameters, 
the masses of the $\zp$-ino and of the MSSM squarks and gauginos,
which are the most robust predictions of this scenario.
At the SUSY breaking scale, $\Lambda_S$,
SUSY breaking in the hidden sector is assumed to generate a SUSY
breaking mass for the fermionic component of the ${\rm U}(1)_{\rm B-L}$ vector superfield.
Given details of the hidden sector, its value could be evaluated via the standard technique
of analytical continuation into superspace \cite{ArkaniHamed:1998kj}.
In particular, the gauge kinetic function of the field strength superfield ${\cal W}_{B-L}^\alpha$
at the SUSY breaking scale is  
\begin{eqnarray}
\label{eqn:zpinomass}
\mathcal{L}_{\tilde{Z}_{B-L}} &=& \int d^2 \theta \left[ \frac{1}{g_{B-L}^2
 } + \beta_{B-L}^{hid} \ln
  \left( \frac{\Lambda_S}{M} \right) \right. \nonumber \\ &+& \left.
 \beta^{vis}_{B-L}\ln \left(
  \frac{\Lambda_S}{M_{\tilde{Z}_{B-L}}} \right) \right] {\cal W}_{B-L}^\alpha {\cal W}_{B-L}^\alpha \;,
\end{eqnarray}
where $M$ is the messenger scale, which we have assumed to be around
the SUSY breaking scale, $M \sim
\Lambda_S$. $\beta_{B-L}^{hid}$ and $\beta^{vis}_{B-L}$ are
$\beta$-functions induced by ${\rm U}(1)_{\rm B-L}$ couplings to hidden and visible
sector fields, respectively.    Using analytical
continuation, we replace $M$
with $M+ \theta^2 F$, where $F$ is the SUSY breaking order
parameter.  We obtain the $\tilde{Z}_{B-L}$ mass as $ M_{\tilde{Z}_{B-L}} \sim g_{B-L}^2
\beta_{B-L}^{hid} F/M$.
We assume that the ${\rm U}(1)_{\rm B-L}$ gauge symmetry is
not broken in the hidden sector.
And we assume some sequestering mechanism so that only the $B-L$ gaugino 
obtains a leading order mass term while the threshold corrections to
the squrks and sleptons are only arisen at the next leading order as similar
to the case of the gaugino mediation, where the $B-L$ gaugino lives in the bulk
in a five dimesional setup while squarks and sleptons are put on the brane.
In such a case, only the $B-L$ gaugino obtains a mass while the scalar masses
receive negligible threshold corrections at the lowest order since they
receive volume suppression.

Since all the chiral superfields in the visible sector are
charged under ${\rm U}(1)_{\rm B-L}$, so all the corresponding scalars receive  soft
mass terms at 1-loop of order
\bea
\label{eqn:scalarmass}
m^2_{\tilde{q}_i} &=& \frac{8}{9} \frac{\alpha_{B-L}}{4 \pi} M_{\tilde{Z}_{B-L}}^2
\ln\left(\frac{\Lambda_S}{M_{\tilde{Z}_{B-L}}} \right),
\nonumber\\
m^2_{\tilde{\ell}_i} &=& 8\, \frac{\alpha_{B-L}}{4 \pi} M_{\tilde{Z}_{B-L}}^2
\ln\left(\frac{\Lambda_S}{M_{\tilde{Z}_{B-L}}} \right),
\eea
where $\alpha_{B-L}=g_{B-L}^2/(4\pi)$ and $ Q^f_{B-L}$ is
the ${\rm U}(1)_{\rm B-L}$ charge of $f$.

The MSSM gaugino masses, however,
can only be generated at 2-loop level since they do not directly couple to the ${\rm U}(1)_{\rm B-L}$,
\bea
\label{eqn:gauginomass}
M_a
&=& 4 c_a\, \frac{\alpha_{B-L}}{4 \pi} \frac{\alpha_a}{4 \pi} M_{\tilde{Z}_{B-L}}
\ln\left(\frac{\Lambda_S}{M_{\tilde{Z}_{B-L}}} \right) \;,
\eea
where $(c_1, c_2, c_3) = (\frac{92}{15}, 4,  \frac{4}{3})$.

Since these gaugino masses are proportional to $c_a$, we expect that the
gluino will typically be lighter than the others at $\mu=M_{\tilde{Z}_{B-L}}$,
so the resultant mass spectra of the gauginos are relatively compressed than the other mediation
mechanisms.

From the discussion above, we see that the gauginos are considerably lighter
than the sfermions. Taking  $M_a \simeq 100$ GeV, we find
\begin{equation}
 M_{\tilde{Z}_{B-L}} \ln\left(\frac{\Lambda_S}{M_{\tilde{Z}_{B-L}}} \right) 
\simeq 10^4 ~\TeV
 \end{equation}
 and
\begin{equation}
{m}_{\tilde{f}} \simeq  10^{-1} M_{\tilde{Z}_{B-L}} \simeq  10^{5}~\GeV.
\end{equation}
Hence, in this scheme of Z-prime mediation, all the sfermion masses become
very heavy at around $10^5$ GeV, while the gauginos are kept at at around the weak
scale, $M_a \simeq 100$ GeV, which can in principle provide a natural candidate of the dark matter.

In our choice of parameters, the gravitino mass is given by
\be
m_{3/2} = \frac{\Lambda_S^2}{\sqrt{3} M_{\rm Pl}} = \{ 24\, {\rm keV},~2.4 \,{\rm MeV},~ 240 \,{\rm MeV}\} \;.
\ee
for $\Lambda_S = \{10^7,\, 10^8, \,10^9\}$ GeV.
Hence the gravity mediation contribution to the gaugino masses is much suppressed,
and is well negligible compared to the Z-prime mediated contribution.

\section{RGEs and its numerical evaluations}
Now we consider the RGEs and analyze
the running of the scalar masses $m_{\Delta_1}^2$ and
$m_{\Delta_2}^2$. The key point for implementing the radiative $B-L$
symmetry breaking is that the scalar potential $V(\Delta_1,\Delta_2)$
receives substantial radiative corrections. In particular, a
negative (mass)$^2$ would trigger the $B-L$ symmetry breaking. 
We argue that the masses of Higgs fields $\Delta_1$ and
$\Delta_2$ run differently in the way that $m^2_{\Delta_1}$ can be
negative whereas $m^2_{\Delta_2}$ remains positive. 
The RGE for the $B-L$ coupling and
mass parameters can be derived from the general results for SUSY
RGEs of Ref. \cite{Martin}.

For the RGEs of the Yukawa couplings, we consider to include the additional
contribution from the the $U(1)_{\rm B-L}$ gauge sector.
\bea
16 \pi^2 \mu \frac{d }{d \mu} Y_A
&=&
\left[
\mbox{\sf MSSM + see-saw}
\right]
+  \delta_{A \nu} Y_\nu f^\dag f 
-2\, a_A \,g_{\rm B-L}^2 \,Y_A \;,
\eea
where $(a_u, a_d, a_\nu, a_e)= (\frac{2}{9},\frac{2}{9},2,2)$, and
the [{\sf MSSM + see-saw}] part of the RGEs can be found in the Appendix.

And the RGE of the Yukawa coupling relevant for the right-handed neutrino mass is written by
\bea
16 \pi^2 \mu \frac{d }{d \mu} f
&=&
4 \, {\rm Tr} [f^\dag f] f + 2\,  (Y_\nu^\dag Y_\nu) f +  2 \, f (Y_\nu^\dag Y_\nu)^T
- 12\, g_{B-L}^2\, f  \;.
\eea
The RGEs of the MSSM gauge couplings are the same as MSSM,
while the RGE of the $U(1)_{\rm B-L}$ gauge coupling is given by
\bea
16 \pi^2 \mu \frac{d g_{B-L}}{d \mu}
&=& b_{B-L} \,g_{B-L}^3 \;,
\eea
where $b_{B-L} = 24$.

For the RGEs of the gaugino masses, it can be written as follows.
\bea
16 \pi^2 \mu \frac{d M_{\tilde{Z}_{B-L}}}{d \mu}
&=& 2 b_{B-L} g_{B-L}^3 M_{\tilde{Z}_{B-L}} \;,
\nonumber\\
16 \pi^2 \mu \frac{d M_a}{d \mu}
&=& 
\left[
\mbox{\sf MSSM + see-saw}
\right]
+\frac{4 c_a g_a^2}{16 \pi^2} g_{B-L}^2 M_{\tilde{Z}_{B-L}} \;,
\eea
where $(c_a) = (\frac{92}{15},4,\frac{4}{3})$.

For the RGEs of the A-terms, it can be written as follows.
\bea
16 \pi^2 \mu \frac{d }{d \mu} \tilde{A}_A
&=&
\left[
\mbox{\sf MSSM + see-saw}
\right]
-2 a_A g_{\rm B-L}^2 (\tilde{A}_A - 2 M_{\tilde{Z}_{B-L}} Y_A ) \;,
\eea
where $\tilde{A}_A = A_A Y_A$.
The RGE of the $A_f$-term can be written as
\bea
16 \pi^2 \mu \frac{d }{d \mu} \tilde{A}_f
&=&
\left(
9\, {\rm Tr} [f^\dag f]
+ 2 \, {\rm Tr} [Y_\nu^\dag Y_\nu] \right) \tilde{A}_f
+ 8\, f \,Y_\nu^\dag \tilde{A}_\nu \;.
\eea
The RGEs of the soft scalar masses are given by
\bea
16 \pi^2 \mu \frac{d m_{\Delta_1}^2}{d \mu}
&=&
2\, {\rm Tr} [f^\dag f] m_{\Delta_1}^2 + 4 \, {\rm Tr} [f^\dag m_N^2 f] 
- 32  g_{B-L}^2 |M_{\tilde{Z}_{B-L}}|^2 \;.
\nonumber\\
16 \pi^2 \mu \frac{d m_{\Delta_2}^2}{d \mu}
&=&
- 32  g_{B-L}^2 |M_{\tilde{Z}_{B-L}}|^2 \;.
\nonumber\\
16 \pi^2 \mu \frac{d m_{\tilde{f}}^2}{d \mu}
&=&
\left[
\mbox{\sf MSSM + see-saw}
\right]
- 8  g_{B-L}^2 (Q_{B-L}^{f})^2 |M_{\tilde{Z}_{B-L}}|^2 \;.
\eea
where $Q_{B-L}^{f}$ is the $B-L$ charge of each chiral multiplet 
$f = Q, U^c, D^c, L$ and
\bea
16 \pi^2 \mu \frac{d m_{\tilde{N}^c}^2}{d \mu}
&=&
\left[
\mbox{\sf MSSM + see-saw}
\right]
+ \left( m_{\tilde{N}^c}^2 f^\dag f + f^\dag f m_{\tilde{N}^c}^2                \right)
\nonumber\\
&+&
2\left(f^\dag  m_{\tilde{N}^c}^2 f + m_{\Delta_1}^2 f^\dag f + \tilde{A}_f^\dag \tilde{A}_f   \right)
- 8  g_{B-L}^2 (Q_{B-L}^{f})^2 |M_{\tilde{Z}_{B-L}}|^2 \;.
\eea
For the RGEs of the $\mu'$-term, it can be written as follows.
\bea
16 \pi^2 \mu \frac{d }{d \mu} \mu'
= ({\rm Tr} [f^\dag f] - 16 g_{B-L}^2) \mu' \;.
\eea
In the numerical analysis we take input all the soft SUSY breaking parameters
to be zero at the SUSY breaking scale, in which the SUSY breaking scale
is varied in the range, $\Lambda_S = 10^7 - 10^9$ GeV,
\be
\tilde{A}_A = 0, ~ m_{\tilde{f}} = 0,~M_a = 0
\ee
and use the following inputs
\be
M_{\tilde{Z}_{B-L}}= 8.7 \times 10^5 \,{\rm GeV}\;, ~ f = 4,\ 5,\ 6, \ 7, ~g_{B-L} = 0.5 \;.
\ee
Note that $\tilde{Z}_{B-L}$ has to be decoupled at the mass scale $M_{\tilde{Z}_{B-L}}$.
Here some comments are in order for the above choices of parameters.
For the large values of the Yukawa coupling ($f$), it blows up before reaching
the GUT scale,  so we have to have a cutoff scale below the GUT scale. 
However, since our motivation to consider a model with $B-L$ gauge symmetry is
to find a relation to the origin of the neutrino masses via the see-saw
mechanism. So, we do not assume a simple SU(5) like GUT picture. 
In fact, if one consider the $B-L$ gauge symmetry, the naive GUT picture would be
broken at an intermediate scale while there is a possibility to realize a grand unification
with an intermediate scale.

\begin{figure}[p]
\begin{center}
%\subfigure[$\Lambda_S = 5 \times 10^7$ GeV]
%{\includegraphics[width=.45\linewidth]{Fig1a.eps}\label{Fig1a}}
\subfigure[$\Lambda_S = 1 \times 10^8$ GeV]
{\includegraphics[width=.6\linewidth]{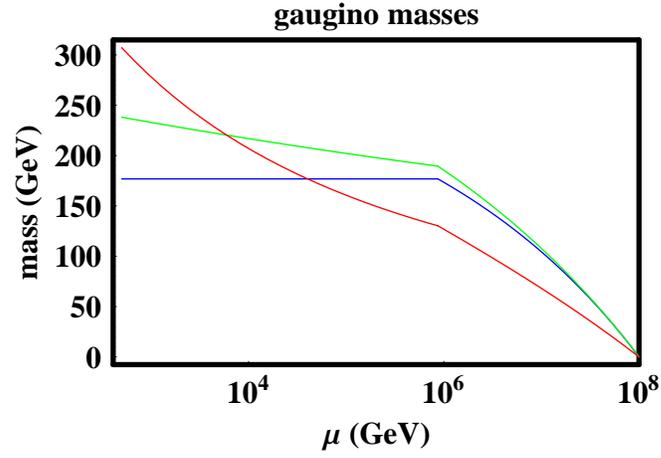}\label{Fig1b}}
\subfigure[$\Lambda_S = 1 \times 10^9$ GeV]
{\includegraphics[width=.6\linewidth]{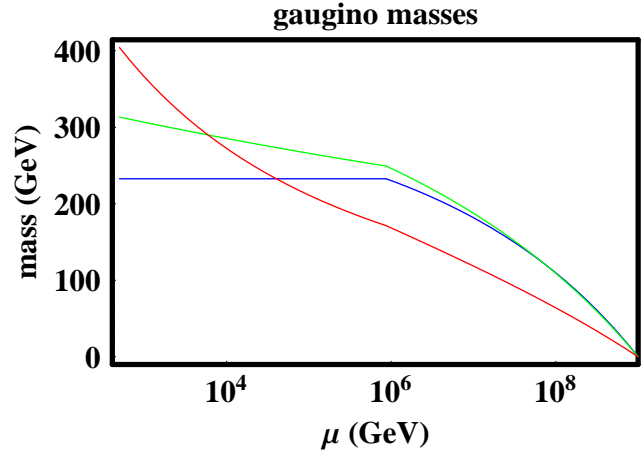}\label{Fig1c}}
\end{center}
\caption{
The evolution of the gaugino masses from the SUSY breaking scale 
to the $B-L$ breaking scale. 
The red line shows the running of the gluino mass, the green line is
the running of the ${\rm SU}(2)$ gaugino mass, and the blue corresponds to the running
of the ${\rm U}(1)_Y$ gaugino mass.
}
\label{Fig1}
\vspace{1cm}
\end{figure}

Using these inputs, in Fig.~{\ref{Fig1}}, we plot the evolution of the gaugino
masses $M_{1,2,3}$ from the SUSY breaking scale to the weak scale. 
In these plots, we varied the SUSY breaking scale as 
$10^8$ GeV and $10^9$ GeV.
It is very interesting that the gluino at the ${\tilde{Z}}_{B-L}$ scale is given as the lightest gaugino,
that is very different from most of the other models of SUSY breaking mediation.
For that reason, the gluino at the weak scale becomes relatively light,
and almost compressed mass spectra for the gaugino sector can be
realized in this scenario, which is very interesting in scope of the LHC.

\begin{figure}[p]
\begin{center}
\includegraphics[width=.8\linewidth]{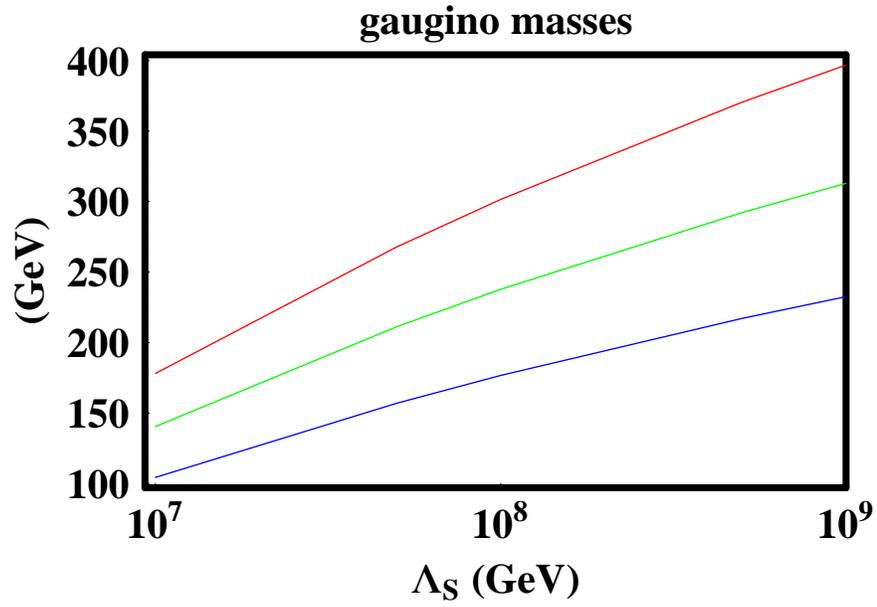}
\end{center}
\caption{
The SUSY breaking scale, $\Lambda_S$, dependence of 
the gaugino masses at the weak scale.
Again, the red line shows the running of the gluino mass, the green line is
the running of the ${\rm SU}(2)$ gaugino mass, and the blue corresponds to the running
of the ${\rm U}(1)_Y$ gaugino mass.
}
\label{Fig2}
\vspace{1cm}
\end{figure}

In Fig.~{\ref{Fig2}}, we plot the SUSY breaking scale, $\Lambda_S$, dependence of 
the gaugino masses $M_{1,2,3}$ at the weak scale.
It simply shows that raising the SUSY breaking scale corresponds to the increase of
the gaugino masses at the weak scale.

\begin{figure}[p]
\begin{center}
\includegraphics[width=.8\linewidth]{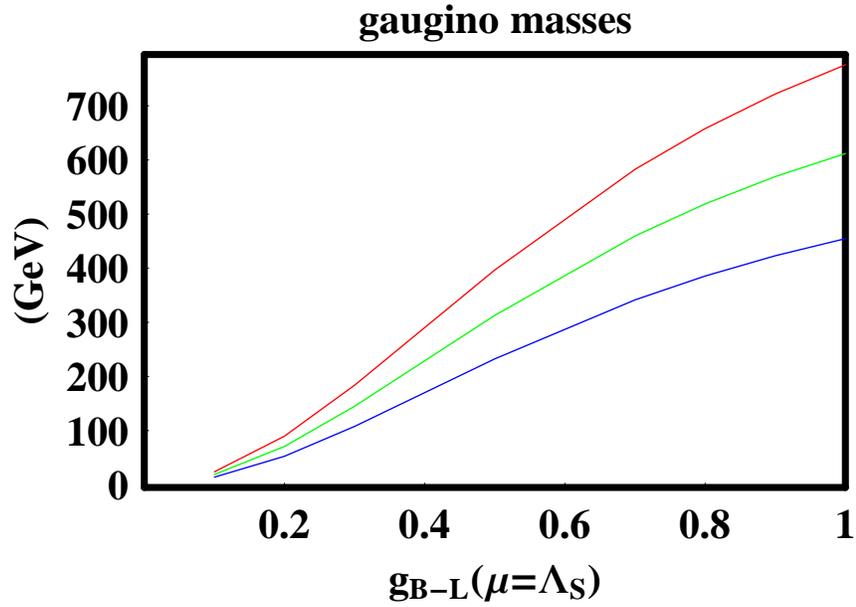}
\end{center}
\caption{
The gauge coupling constant, $g_{\rm B-L}$, dependence of 
the gaugino masses at the weak scale.
Here the gauge coupling constant, $g_{\rm B-L}$ is given at a given SUSY breaking scale
$\Lambda_S = 10^9$ GeV.
Again, the red line shows the running of the gluino mass, the green line is
the running of the ${\rm SU}(2)$ gaugino mass, and the blue corresponds to the running
of the ${\rm U}(1)_Y$ gaugino mass.
}
\label{Fig3}
\vspace{1cm}
\end{figure}

In Fig.~{\ref{Fig3}}, we plot the gauge coupling constant, $g_{\rm B-L}$, dependence of 
the gaugino masses at the weak scale.
Here the gauge coupling constant, $g_{\rm B-L}$ is given at a given SUSY breaking scale
$\Lambda_S = 10^9$ GeV.

\begin{figure}[p]
\begin{center}
\includegraphics[width=.8\linewidth]{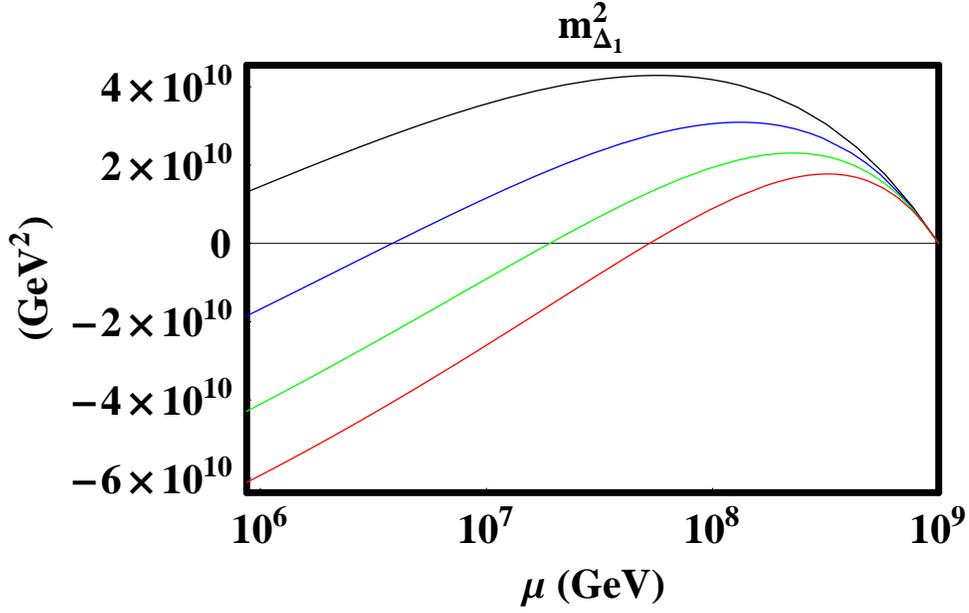}
\end{center}
\caption{
The evolution of the soft mass squared for the field $\Delta_1$
from the SUSY breaking scale to the $B-L$ breaking scale.
In this plot, we take the SUSY breaking scale as $\Lambda_S = 10^9$ GeV,
$M_{\tilde{Z}_{B-L}}= 8.7 \times 10^5 \,{\rm GeV}$ and $g_{B-L} = 0.5$.
In this figure, from top to the bottom curves, we varied the value of $f$ as $f=4,\ 5,\ 6,\ 7$.
}
\label{Fig4}
\vspace{1cm}
\end{figure}

\begin{figure}[p]
\begin{center}
\includegraphics[width=.8\linewidth]{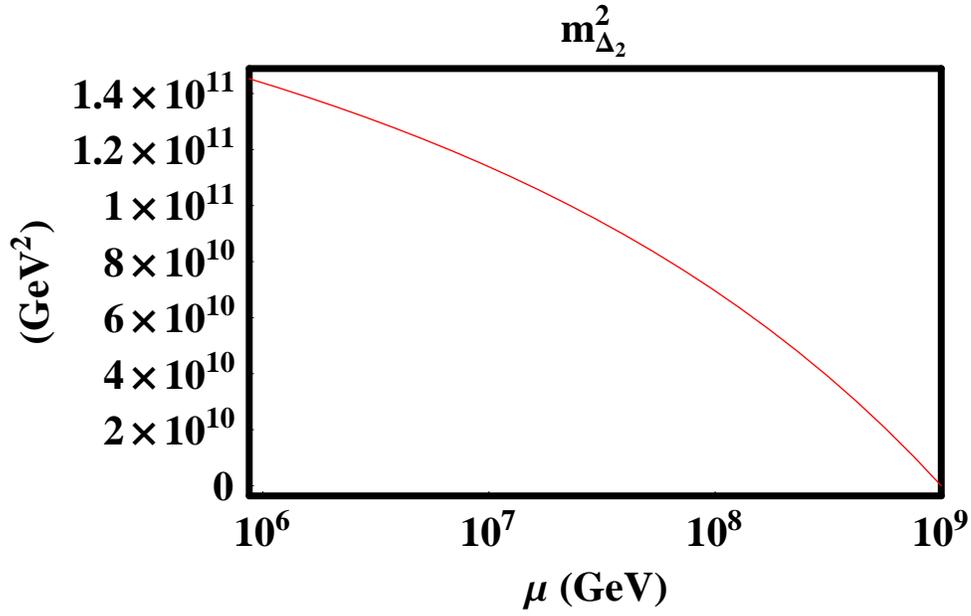}
\end{center}
\caption{
The evolution of the soft mass squared for the field $\Delta_2$
from the SUSY breaking scale to the $B-L$ breaking scale.
In this plot, we take the SUSY breaking scale as $\Lambda_S = 10^9$ GeV,
$M_{\tilde{Z}_{B-L}}= 8.7 \times 10^5 \,{\rm GeV}$ and $g_{B-L} = 0.5$.
}
\label{Fig5}
\vspace{1cm}
\end{figure}

The evolutions of the soft mass squared for the fields $\Delta_1$ and $\Delta_2$
are plotted in Fig.~{\ref{Fig4}} and Fig.~{\ref{Fig5}} for a given SUSY breaking scale
as $\Lambda_S = 10^9$ GeV. 
%In these figures, it is definitely seen the physical threshold at the scale, $M_{\tilde{Z}_{B-L}}$ 
%since we have decoupled $\tilde{Z}_{B-L}$ at this scale.
In Fig.~{\ref{Fig4}}, from top to the bottom curves, we varied the value of $f$ as $f=4,\, 5,\, 6,\, 7$.
For example, for the case of $f=5$, the soft mass squared for the fields $\Delta_1$
goes across the zeros at the scale about $10^6$ GeV toward negative value, that is nothing but the realization
of the radiative symmetry breaking of ${\rm U}(1)_{B-L}$ gauge symmetry.
The running behavior in Fig.~{\ref{Fig4}} can be understood in the following way.
At first, starting from the high energy scale, the soft mass squared increases
because of the gauge coupling contributions, and decrease of the mass squared
is caused by the Yukawa coupling that dominate over the gauge coupling contribution
at some scale. 
Next, since at the mass scale of $\tilde{Z}_{B-L}$, it is decoupled from the RGEs, 
there are only the Yukawa coupling contributions to
the soft mass squared which rapidly decreases to across the zeros.
Therefore, the radiative $B-L$ symmetry breaking can naturally be realized.

The see-saw scale, which is found to be at $v_{B-L} = 10^5$ GeV, hence
the right-handed neutrino obtains a mass of $M_N = f v_{B-L} = 5 \times 10^5$ GeV.
This scale of the right-handed neutrino is nice for the thermal leptogenesis to be viable
in supersymmetric models with gravity mediation.

%\newpage

\section{Summary}
We have shown that a mechanism of radiative $B-L$ symmetry breaking
can work in analogous to the RESB.
The breaking scale of $B-L$ symmetry is related to the neutrino masses
through the see-saw mechanism.
Once we incorporate the ${\rm U}(1)_{\rm B-L}$ gauge symmetry in SUSY models,
the ${\rm U}(1)_{\rm B-L}$ gaugino, $\tilde{Z}_{B-L}$ can provide all the soft masses
in the MSSM.
Then we find a link between the neutrino mass (more precisly the see-saw or $B-L$ scale
of order $10^{5}$ GeV) and the Z-prime mediated SUSY breaking scale.
In this scheme of Z-prime mediation, all the sfermion masses become
very heavy at around $10^5$ GeV, while the gauginos are kept at at around the weak
scale, $M_a \simeq 100$ GeV.
It is also very interesting that the gluino at $\tilde{Z}_{B-L}$ scale is given as the lightest gaugino,
that is very different from most of the other models of SUSY breaking mediation.
For that reason, the gluino at the weak scale becomes relatively light,
and almost compressed mass spectra for the gaugino sector can be
realized in this scenario, which is very interesting in scope of the LHC.

\section*{Acknowledgments}
T. Kikuchi would like to thank K.S. Babu
 for his hospitality at Oklahoma State University.
The work of T.Kikuchi is supported by the Research
Fellowship of the Japan Society for the Promotion of Science (\#1911329).

%\newpage

\begin{appendix}
\section{RGEs in the MSSM with right-handed neutrinos}
%%%%%%%%%%%%%%%%%%%%%%%%%%%%%%%%%%%%%%%%%%%%%%%%%%%%%%%%%%%
\subsection{The 2-loop RGE for the gauge couplings}
%%%%%%%%%%%%%%%%%%%%%%%%%%%%%%%%%%%%%%%%%%%%%%%%%%%%%%%%%%%
\bea
16\pi^2 \mu \frac{d}{d \mu} ~g_1 &\!=\!& \frac{33}{5}\, g_1^3 
+ \frac{g_1^3}{16\pi^2} \left(\frac{199}{25} g_1^2 + \frac{27}{5} g^2_2
+ \frac{88}{5} g_3^2\right) \nn, \\
16\pi^2 \mu \frac{d}{d \mu} ~g_2 &\!=\!& \, g_2^3 + 
\frac{g_2^3}{16\pi^2} \left(\frac{9}{5} g_1^2 + 25 g^2_2
+ 24 g_3^2\right) \nn, \\
16\pi^2 \mu \frac{d}{d \mu} ~g_3 &\!=\!& -3 \, g_3^3 
+ \frac{g_3^3}{16\pi^2} \left(\frac{1}{5} g_1^2 + 9 g^2_2
+ 14 g_3^2\right) \nn.
\eea
Here $g_2 \equiv g$ is the $SU(2)_L$ gauge coupling constant and 
$g_1 \equiv \sqrt{\frac{5}{3}} g^\prime$ is the $U(1)$ gauge coupling 
constant with the GUT normalization 
($g_1 = g_2 = g_3$ at $\mu = M_{\rm GUT}$). 

%%%%%%%%%%%%%%%%%%%%%%%%%%%%%%%%%%%%%%%%%%%%%%%%%%%%%%%%%%
\subsection{The 1-loop RGE for the Yukawa couplings}
%%%%%%%%%%%%%%%%%%%%%%%%%%%%%%%%%%%%%%%%%%%%%%%%%%%%%%%%%%
\bea
16\pi^2 \mu \frac{d}{d \mu} Y_{u} &\!=\!& 
Y_u \left[
\left \{ -\frac{13}{15} g_1^2 - 3 g_2^2 -\frac{16}{3} g_3^2 
  + 3 \,{\rm Tr} ( Y_u^{\dagger} Y_u )
  +   {\rm Tr} ( Y_\nu^{\dagger} Y_\nu ) \right \}
{\bf 1}_{3\times3}
\right.
\nonumber \\
&& \left.\; + 3 \, ( Y_u^{\dagger} Y_u) 
      + (Y_{d}^{\dagger} Y_{d}) \right]\nn, \\
16\pi^2 \mu \frac{d}{d \mu} Y_{d} &\!=\!& 
Y_d \left[
\left \{ -\frac{7}{15} g_1^2 - 3 g_2^2 -\frac{16}{3} g_3^2 
  + 3 \,{\rm Tr} ( Y_d^{\dagger} Y_d )
  +   {\rm Tr} ( Y_e^{\dagger} Y_e ) 
\right \} {\bf 1}_{3\times3}
\right.
\nonumber \\
&& \left.
+ \; 3 \, ( Y_d^{\dagger} Y_d) 
+ (Y_{u}^{\dagger} Y_{u}) \right]\nn, \\
16\pi^2 \mu \frac{d}{d \mu} Y_{\nu} &\!=\!&
Y_\nu \left[
\left \{ -\frac{3}{5} g_1^2 - 3 g_2^2 
+ 3 \,{\rm Tr} \left( Y_{u}^{\dagger} Y_{u} \right)
+   {\rm Tr} \left( Y_{\nu}^{\dagger} Y_{\nu} \right) 
\right \}  {\bf 1}_{3\times3}
\right.
\nonumber \\
&& \left.
+ 3 \, \left( Y_{\nu}^{\dagger} Y_{\nu} \right)
+ \left(Y_{e}^{\dagger} Y_{e} \right) \right] 
\nn, \\
16\pi^2 \mu \frac{d}{d \mu} Y_{e} &\!=\!& 
Y_e \left[
\left \{ -\frac{9}{5} g_1^2 - 3 g_2^2 
  + 3 \,{\rm Tr} ( Y_d^{\dagger} Y_d)
  +   {\rm Tr} ( Y_e^{\dagger} Y_e) 
\right \} {\bf 1}_{3\times3}
\right.
\nonumber \\
&& \left.
+ 3 \, \left( Y_e^{\dagger} Y_e \right) + 
\left(Y_{\nu}^{\dagger} Y_{\nu} \right) \right]\nn. 
\eea
%

%%%%%%%%%%%%%%%%%%%%%%%%%%%%%%%%%%%%%%%%%%%%%%%%%%%%%%%%%%%
% The 1-loop RGE for the soft SUSY breaking terms:
%%%%%%%%%%%%%%%%%%%%%%%%%%%%%%%%%%%%%%%%%%%%%%%%%%%%%%%%%%%%%
\subsection{The 2-loop RGE for the gaugino masses}
%%%%%%%%%%%%%%%%%%%%%%%%%%%%%%%%%%%%%%%%%%%%%%%%%%%%%%%%%%%%
\bea
16\pi^2 \mu \frac{d}{d \mu} M_1 &\!=\!& \frac{66}{5} \, g_1^2 M_1 
\nonumber\\
&\!+\!&
\frac{2 g_1^2}{16\pi^2} \left\{ \frac{199}{5} g^2_1 \left( 2 M_1  
\right)  + \frac{27}{5} g_2^2 \left( M_1 + M_2 \right) 
+ \frac{88}{5} g_3^2 \left( M_1 + M_3 \right)\right\} \nn, \\
16\pi^2 \mu \frac{d}{d \mu} M_2 &\!=\!& 2 \, g_2^2 M_2 
\nonumber\\
&\!+\!&
\frac{2 g_2^2}{16\pi^2} \left\{ \frac{9}{5} g^2_1 \left(M_1 + M_2 \right)
+ 25 g_2^2 \left( 2 M_2\right) + 24 g_3^2 
\left( M_2 + M_3 \right) \right\} \nn, \\
16\pi^2 \mu \frac{d}{d \mu} M_3 &\!=\!& -6\, g_3^2 M_3 
\nonumber\\
&\!+\!&
\frac{2 g_3^2}{16\pi^2} \left\{ \frac{11}{5} g^2_1 \left(M_1 + M_3 \right)
+ 9 g_2^2 \left( M_2 + M_3 \right) + 14 g_3^2 
\left( 2 M_3 \right) \right\} \nn.
\eea
%%%%%%%%%%%%%%%%%%%%%%%%%%%%%%%%%%%%%%%%%%%%%%%%%%%%%%%%%%%%
\subsection{The 1-loop RGE for the soft SUSY breaking mass terms}
%%%%%%%%%%%%%%%%%%%%%%%%%%%%%%%%%%%%%%%%%%%%%%%%%%%%%%%%%%%%%
\bea
16\pi^2 \mu \frac{d}{d \mu} \left( m^2_{\tilde{\sss q}} \right)_{ij} 
&\!=\!&
- \left( \frac{2}{15} g_1^2 \left| M_1 \right|^2 
+ 6 g_2^2 \left| M_2 \right|^2 + \frac{32}{3} g_3^2 \left| M_3 
\right|^2\right) \delta_{ij}
+ \frac{1}{5} g_1^2~S~\delta_{ij} \nonumber \\
&\!+\!& 
\left( m^2_{\tilde{\sss q}} Y_u^{\dagger} Y_u 
+ m^2_{\tilde{\sss q}} Y_d^{\dagger} Y_d 
+ Y_u^{\dagger} Y_u m^2_{\tilde{\sss q}} 
+ Y_d^{\dagger} Y_d  m^2_{\tilde{\sss q}} \right)_{ij} \nonumber \\
&\!+\!&
2 \left( Y_u^{\dagger} m^2_{\tilde u} Y_u
           + {m}^2_{H_u} Y_u^{\dagger} Y_u
+ A_u^{\dagger} A_u \right)_{ij} \nonumber \\
&\!+\!&
2 \left( Y_d^{\dagger} m^2_{\tilde {d}} Y_{d}
+ {m}^2_{H_d} Y_{d}^{\dagger} Y_{d}
+ A_{d}^{\dagger} A_{d} \right)_{ij}\nn, \\
%%%%%%%%%%%%%%%%%%%%%%%%%%%%%%%%%%%%%%%%%%%%%%%%%%%%%%%
16\pi^2 \mu \frac{d}{d \mu} \left( m^2_{\tilde{u}} \right)_{ij} 
&\!=\!&
- \left( \frac{32}{15} g_1^2 \left| M_1 \right|^2 
+ \frac{32}{3} g_3^2 \left| M_3 \right|^2\right) \delta_{ij}
- \frac{4}{5} g_1^2~S~\delta_{ij} \nonumber \\
&\!+\!&
 2 \left( m^2_{\tilde u} 
Y_u^{\dagger} Y_u + Y_u^{\dagger} Y_u m^2_{\tilde u} \right)_{ij} 
\nonumber \\
&\!+\!&
 4 \left( Y_u m^2_{\tilde{\sss q}} Y_u^{\dagger} + {m}^2_{H_u} 
Y_u^{\dagger} Y_u + A_u A_u^{\dagger} \right)_{ij}\nn, \\
16\pi^2 \mu \frac{d}{d \mu} \left( m^2_{\tilde{d}} \right)_{ij} 
&\!=\!&
- \left( \frac{8}{15} g_1^2 \left| M_1 \right|^2 
+ \frac{32}{3} g_3^2 \left| M_3 \right|^2\right) \delta_{ij}
+ \frac{2}{5} g_1^2~S~\delta_{ij} \nonumber \\
&\!+\!& 
2 \left( m^2_{\tilde d} 
Y_d^{\dagger} Y_d + Y_d^{\dagger} Y_d m^2_{\tilde d} \right)_{ij} 
\nonumber \\
&\!+\!&
4 \left( Y_d m^2_{\tilde{\sss q}} Y_d^{\dagger} + {m}^2_{H_d} 
Y_d^{\dagger} Y_d + A_d A_d^{\dagger} \right)_{ij}\nn, \\
%
%%%%%%%%%%%%%%%%%%%%%%%%%%%%%%%%%%%%%%%%%%%%%%%%%%%%%%%
%
16\pi^2 \mu \frac{d}{d \mu} \left( m^2_{\tilde{\sss \ell}} 
\right)_{ij}&\!=\!&   -\left( \frac{6}{5} g_1^2 \left| M_1 \right|^2 
+ 6 g_2^2 \left| M_2 \right|^2 \right) \delta_{ij}
-\frac{3}{5} g_1^2~S~\delta_{ij} \nonumber \\
&\!+\!&
\left( m^2_{\tilde{\sss \ell}} Y_e^{\dagger} Y_e 
+ m^2_{\tilde{\sss \ell}} Y_\nu^{\dagger} Y_\nu 
+ Y_e^{\dagger} Y_e m^2_{\tilde{\sss \ell}} 
+ Y_\nu^{\dagger} Y_\nu  m^2_{\tilde{\sss \ell}} \right)_{ij} \nonumber \\
&\!+\!&
2 \left( Y_e^{\dagger} m^2_{\tilde e} Y_e
           +{m}^2_{H_d} Y_e^{\dagger} Y_e
+ A_e^{\dagger} A_e \right)_{ij} \nonumber \\
&\!+\!&
2 \left( Y_{\nu}^{\dagger} m^2_{\tilde {\sss\nu}} Y_{\nu}
+ {m}^2_{H_u} Y_{\nu}^{\dagger} Y_{\nu}
+ A_{\nu}^{\dagger} A_{\nu} \right)_{ij}\nn, \\
16\pi^2 \mu \frac{d}{d \mu} \left( m^2_{\tilde{e}} \right)_{ij}&\!=\!& 
- \frac{24}{5} g_1^2 \left| M_1 \right|^2 \delta_{ij}
+ \frac{6}{5} g_1^2~S~\delta_{ij} + 2 \left( m^2_{\tilde e} 
Y_e^{\dagger} Y_e + Y_e^{\dagger} Y_e m^2_{\tilde e} \right)_{ij} 
\nonumber \\
&\!+\!&
4 \left( Y_e m^2_{\sss\widetilde \ell} Y_e^{\dagger} + {m}^2_{H_d} 
Y_e^{\dagger} Y_e + A_e A_e^{\dagger}\right)_{ij}\nn, \\
16\pi^2 \mu \frac{d}{d \mu} \left( m^2_{\tilde{\nu}} \right)_{ij} &\!=\!& 
2 \left( m^2_{\tilde \nu} Y_{\nu}^{\dagger} Y_{\nu} 
+ Y_{\nu}^{\dagger} Y_{\nu} m^2_{\tilde \nu} \right)_{ij} 
+ 4 \left( Y_{\nu} m^2_{\tilde \ell} Y_{\nu}^{\dagger}
+ {m}^2_{H_u} Y_{\nu}^{\dagger} Y_{\nu}
+ A_{\nu} A_{\nu}^{\dagger}\right)_{ij} \nn. 
\eea
\bea
{16 \pi^2} \mu \frac{d}{d \mu} (m^2_{H_u}) &=&  -\left(\frac{6}{5} 
g^{2}_{1} \abs{M_{1}}^{2} + 6 g^{2}_{2} \abs{M_{2}}^{2}\right)
+ \frac{3}{5} g_1^2 S
\nonumber\\
&\!+\!& 
6\, {\rm {\rm Tr}}\left(
m^2_{\tilde q} Y^{\dagger}_{u} Y_{u} + Y^{\dagger}_{u}
( m^2_{\tilde u} + m^2_{H_u} ) Y_{u} + A^{\dagger}_{u} A_{u}
\right) \nonumber\\
&\!+\!& 2\, {\rm Tr} \left( m^2_{\tilde{\sss \ell}} Y^{\dagger}_{\nu} Y_{\nu}
+ Y^{\dagger}_{\nu} ( m^2_{\tilde \nu} + m^2_{H_u}) Y_{\nu} 
+ A^{\dagger}_{\nu} A_{\nu} \right) \nn,\\
%Hd=h_d
{16 \pi^2} \mu \frac{d}{d \mu} (m^2_{H_d}) &=& -\left( \frac{6}{5} 
g^{2}_{1} \abs{M_{1}}^{2} + 6 g^{2}_{2} \abs{M_{2}}^{2}\right) 
- \frac{3}{5} g_1^2 S \nonumber\\
&\!+\!&
6\, {\rm Tr}\left( m^2_{\tilde q} Y^{\dagger}_{d} Y_{d} 
+ Y^{\dagger}_{d} 
( m^2_{\tilde d} + m^2_{H_d}) Y_{d} + A^{\dagger}_{d} A_{d}\right)
\nonumber\\
&\!+\!& 2\, {\rm Tr} \left( m^2_{\tilde{\sss \ell}} Y^{\dagger}_{e} Y_{e} 
+ Y^{\dagger}_{e} ( m^2_{\tilde e} + m^2_{H_d}) Y_{e} 
+ A_{e}^{\dagger} A_{e} \right) \nn,
\eea
where
\begin{equation}
S \equiv {\rm Tr} (m_{\tilde q}^2 + m_{\tilde d}^2 - 2 m_{\tilde u}^2
- m_{\tilde \ell}^2 + m_{\tilde e}^2 ) - {m}^2_{H_d} 
+ {m}^2_{H_u} \nn.
\end{equation}
%%%%%%%%%%%%%%%%%%%%%%%%%%%%%%%%%%%%%%%%%%%%%%%%%%%%%%%%%%%%
\subsection{The 1-loop RGE for the soft SUSY breaking A-terms}
%%%%%%%%%%%%%%%%%%%%%%%%%%%%%%%%%%%%%%%%%%%%%%%%%%%%%%%%%%%%%
\bea
16\pi^2 \mu \frac{d}{d \mu} A_{u_{ij}}&=&
\left\{ -\frac{13}{15} g_1^2 - 3 g_2^2 - \frac{16}{3} g_3^2 + 3 {\rm Tr} 
(Y_u^{\dagger} Y_u) + {\rm Tr}( Y_{\nu}^{\dagger} Y_{\nu}) 
\right \} A_{u_{ij}} 
\nonumber \\
&\!+\!&
2 \left\{ \frac{13}{15} g_1^2 M_1 + 3 g_2^2 M_2 +\frac{16}{3} g_3^2 M_3
+ 3 {\rm Tr}(Y_u^{\dagger} A_u)
+ {\rm Tr}( Y_{\nu}^{\dagger} A_{\nu}) \right \} Y_{u_{ij}}
\nonumber \\
&\!+\!& 
4 ( Y_u^{\dagger} Y_u A_u)_{ij} 
+ 5 ( A_u Y_u^{\dagger} Y_u)_{ij} 
+ 2 ( Y_u Y_d^{\dagger} A_d)_{ij} 
+ ( A_u Y_d^{\dagger} Y_d)_{ij} \nn,\\
%Ad
16 \pi^2 \mu \frac{d}{d \mu} A_{d_{ij}}&=&
\left\{
-\frac{7}{15} g_1^2 - 3 g_2^2  -\frac{16}{3} g_3^2 
+ 3 {\rm Tr} ( Y_d^{\dagger} Y_d)
+ {\rm Tr}( Y_{e}^{\dagger} Y_{e}) \right \} A_{d_{ij}}
\nonumber \\
&\!+\!& 2 \left\{ \frac{7}{15} g_1^2 M_1 + 3 g_2^2 M_2 + 
\frac{16}{3} g_3^2 M_3 + 3 {\rm Tr}( Y_d^{\dagger} A_d) 
+ {\rm Tr} ( Y_{e}^{\dagger} A_{e}) \right \} Y_{d_{ij}}
\nonumber \\ 
&\!+\!&
4 ( Y_d^{\dagger} Y_d A_d)_{ij} 
+ 5 ( A_d Y_d^{\dagger} Y_d)_{ij} 
+ 2 ( Y_d Y_u^{\dagger} A_u)_{ij} 
+ ( A_d Y_u^{\dagger} Y_u)_{ij} \nn, 
\\
16\pi^2 \mu \frac{d}{d \mu} A_{e_{ij}}&\!=\!&  
 \left\{ -\frac{9}{5} g_1^2 -3 g_2^2
+ 3 {\rm Tr} ( Y_d^{\dagger} Y_d )
+   {\rm Tr} ( Y_e^{\dagger} Y_e ) \right \} A_{e_{ij}} \nonumber \\
&\!+\!&
2 \left\{
\frac{9}{5} g_1^2 M_1 + 3 g_2^2 M_2 
+ 3 {\rm Tr} ( Y_d^{\dagger} A_d)
+   {\rm Tr} ( Y_e^{\dagger} A_e) \right \} Y_{e_{ij}} \nonumber \\
&\!+\!&
4 \left( Y_e^{\dagger} Y_e A_e \right)_{ij} 
+ 5 \left( A_e Y_e^{\dagger} Y_e \right)_{ij} 
+ 2 \left( Y_e Y_{\nu}^{\dagger} A_{\nu} \right)_{ij} 
+  \left( A_e Y_{\nu}^{\dagger} Y_{\nu} \right)_{ij} \nn, 
\\
16\pi^2 \mu \frac{d}{d \mu} A_{\nu_{ij}}&\!=\!&  
\left\{ -\frac{3}{5} g_1^2 -3 g_2^2 
+ 3 {\rm Tr} ( Y_u^{\dagger} Y_u)
+  {\rm Tr} ( Y_{\nu}^{\dagger} Y_{\nu}) \right \} A_{\nu_{ij}} \nonumber \\
&\!+\!&
2 \left\{ \frac{3}{5} g_1^2 M_1 + 3 g_2^2 M_2 
+ 3 {\rm Tr} ( Y_u^{\dagger} A_u)
+   {\rm Tr} ( Y_{\nu}^{\dagger} A_{\nu}) \right \} Y_{\nu_{ij}} 
\nonumber \\
&\!+\!&
 4 ( Y_{\nu}^{\dagger} Y_{\nu} A_{\nu})_{ij}
+ 5 ( A_{\nu} Y_{\nu}^{\dagger} Y_{\nu})_{ij} 
+ 2 ( Y_{\nu} Y_e^{\dagger} A_e)_{ij}
+ ( A_{\nu} Y_e^{\dagger} Y_e)_{ij} \nn. 
\eea

\end{appendix}

%\newpage

%%%%%%%%%%%%%%%%%%%%%%%%%%%%%%%%%%%%%%

%
\pagestyle{empty}
%%%%%%%%%%%%%%%%%%%%%%%%%%%%%%%%%
%    Figures
%%%%%%%%%%%%%%%%%%%%%%%%%%%%%%%%%
%%%%%%%%%%%%%%%%%%%%%%%%%%%%%%%%%
%%%%%%%%%%%%%%%%%%%%%%%%%%%%%%%%%
%%%%%%%%%%%%%%%%%%%%%%%%%%%%%%%%%
%
%%%%%%%%%%%%%%%%%%%%%%%%%%%%%%%%%
\end{document}